\documentclass[11pt,fleqn]{article}

\usepackage{amsmath}
\usepackage{amsfonts}
\usepackage{amssymb}

\begin{document}

\Large

\begin{center}
Extension of Realisations for Low-Dimensional \\
Lie Algebras and Relative Differential Invariants

\hspace{10pt}

% Author names and affiliations
\large
Iryna Yehorchenko

\small
Institute of Mathematics, NAS of Ukraine, Ukraine\\
iyegorch@imath.kiev.ua                                \\
Institute of Mathematics, PAN, Poland\\
iyehorchenko@impan.pl
\end{center}

\hspace{10pt}

\footnotesize

{\bf Abstract}.

We consider realisations of Lie algebras $L=\langle Q_m\rangle$ with basis operators
\begin{equation}\nonumber
 Q_m=\zeta_{ms}({x_i})\partial_{x_s},
\end{equation}
\noindent
where ${x_i}$ are some variables that may be regarded as dependent or independent in construction of some equations or differential invariants, and $\zeta_{ms}$ are some functions of ${x_i}$.

We take additional variables $R_k$, and study
the linearly extended action operators
$ {\hat Q_m}={Q_m}+\lambda_{mjk}R_j\partial_{R_k}$
that form the same Lie algebra with the same structural constants. For a fixed realisation of any Lie algebra $L$ we can classify all inequivalent extended action realisations for a finite number of additional variables. Such realisations allow to construct new invariant equations for the same realisation of the algebra, but involving additional variables, and to classify exhaustively relative differential invariants and invariant equations for the respective realisations of Lie algebras. They can be also applied to other problems in the symmetry analysis of differential equations. Here we classify extensions of realisations for inequivalent low-dimensional Lie algebras.

The problem of classification of the relative differential invariants has an interesting story attached to it, and the background section may be worth reading not only for people in the field.
\newpage
\normalsize

\section{Background}

We consider construction of extensions for realisations of low-dimensional Lie algebras that were listed and classified in \cite{{IYe:Maryna},{IYe:PBNL2003}}. Here we will consider only one- and two-dimensional algebras, and the three-dimensional Poincar\'e algebra for one space variable with the aim to present main ideas.

We start with a realisation of a Lie algebra with the basis operators $L=\langle Q_m\rangle$,
\begin{equation}\label{IYe:fodo}
 Q_m=\zeta_{ms}({x_i})\partial_{x_s},
\end{equation}
\noindent
where ${x_i}$ are some variables that may be regarded as dependent or independent in construction of some equations or differential invariants, and $\zeta_{ms}$ are some functions of ${x_i}$. In the examples there dependent variables are designated specifically, we may use other letters. Hereinafter we will imply summation over the repeated indices, if not specially indicated otherwise.

We take additional variables $R_k$, and study
extended action operators $ {\hat Q_m}={Q_m}+\lambda_{mjk}R_j\partial_{R_k}$
that form the same Lie algebra with the same structural constants. For a realisation of any Lie algebra $L$, we can classify all inequivalent extended action realisations for a finite number of additional variables. That allows constructing new invariant equations for the same realisation of the algebra, but involving additional variables, and to classify exhaustively relative differential invariants (RDIs) and invariant equations for the respective realisations of Lie algebras.

The problem of classification of the extended realisations (not only linear) is interesting by itself, but for classification of relative invariants of Lie algebras we need specifically only linear extensions with nonzero coefficients at $R\partial_R$.

Relative differential invariants (not only absolute) are needed, in particular, for full group classification of PDE invariant under some Lie algebras. Classification of relative differential invariants is an interesting problem of the abstract algebra, even without any relevance to differential equations and their symmetry.

{\bf Definition 1}. A function $\Theta$ depending on $x,u$ and on partial derivatives $u$ of order up to $l$ ($\Theta$ may be a set of functions
$(\Theta_1,....,\Theta_N)$) is called an RDI for the Lie algebra $L=\langle Q_m\rangle$, if it is an invariant
of the $l$-th Lie prolongation of this algebra:
$$
\mathop{Q}\limits^l\!{}_m
\Theta(x,u,\mathop{u}\limits_1,\ldots,\mathop{u}\limits_l)= \lambda_m(x,u,\mathop{u}\limits_1,
\ldots,\mathop{u}\limits_l)\Theta,
$$
\noindent where $\lambda_m$ are some functions; if $\lambda_m=0$,
$\Theta$ is an absolute differential invariant (ADI) of the algebra
$L$; if $\lambda_m \ne 0$, it is a proper RDI.

{\bf Definition 2}.
A maximal set of functionally independent invariants of the
order $r\leq l$ of a Lie algebra $L$ is a functional basis of differential invariants of the order $l$ for the algebra $L$.

Note that we cannot treat a set of independent RDIs the same way as we would treat a set of ADIs -- a function of RDIs may be not invariant. E.g. linear combinations of RDIs generally speaking will not be RDIs.

This paper will come with a story.
The methods for full description and classification of absolute differential invariants are well-known. Some mathematicians (e.g. Oliver Glenn in \cite{IYe:Glenn1928}) claimed that they know how to classify the RDI, but I failed to find their papers with such description. In the cited paper Oliver Glenn wrote in section VII "Remarks concerning relative invariants" that he originated a theory of relative differential invariants by employing the method of invariant elements. He gives a reference \cite{IYe:Glenn1924}, but that reference does not contain anything relevant to classification of relative differential invariants for arbitrary Lie algebras.

Description and classification of relative invariants of groups and algebras may be used for full description of invariant differential equations (absolute differential invariants alone do not provide a full description of such equations), and is of interest as an independent algebraic problem.

I was inspired to solve this problem by the paper by Mark Fels and Peter Olver \cite{IYe:F-O}, who described the need for such classification, but did not present it. I saw that their hints to solution did not work, and found my own way  how to do such classification.

However, some time after I presented my solution at the SPT-2007 conference in Otranto, Pavel Winternitz who was present at my talk sent me their paper with Jos\'e Cari\~{n}ena and Mariano del Olmo, published in a collection of papers of the same Group Symposium in Moscow that was held in 1990 (\cite{IYe:C-DO-W}, and also see their later paper \cite{IYe:C-DO-W93}) .

I was not present at the talk by Pavel Winternitz then in 1990, as it was a parallel session talk scheduled at exactly the same time as the talk of my scientific adviser Wilhelm Fushchych. Maybe it was just accidental, maybe the organisers knew about some sort of conflict between them because of priority of establishment of the concept of the conditional symmetry and did not want any heated discussions at the session.

Further publication of that Symposium papers was seriously disrupted by political events of the Soviet Union destruction. It cost us (mostly me) immense efforts worth a special story, to prepare our papers in the required format on computers we did not have then (the price of a needed personal computer then was equivalent to a price of a new car or an apartment in Kyiv), and to get the $5^{\prime\prime}$
 floppy  disks to Moscow to the conference organizers in early 1991 (no email was available, the disks were to be transported by train, and train tickets also were not available without special efforts). Some time later we were told by the conference organisers that our papers cannot be published because there are no more money left due to huge inflation in the then yet Soviet Union.

We did not find out whether our papers (and papers of other conference participants) were actually published at all until around 2010 (almost 20 years later) when Google started scanning and indexation of conference proceedings. Part of the conference papers were published in a volume of Lecture Notes (that was not available to us at all, and was not scanned by Google for copyright reasons), and the remainder, our papers including, was published in another volume of just conference proceedings with less stringent copyright rules.

So - both Peter Olver and Mark Fels, and I did not know that the problem of classification of the relative differential invariants was actually solved in 1990 under a different name of the problem and of the solution. Both solutions (the Winternitz and co-authors' version using cohomology of the second type and my version using extensions of realisations of Lie algebras) provide equivalent results.

The method for RDI classification I presented in Otranto in 2007 was published in the conference proceedings (Symmetry and Perturbation Theory: Proceedings of the International Conference SPT 2007, Otranto Italy, 2-9 June 2007) \cite{IYe:IYeProcSPT2007}.

In the case when the task is to construct general extension operators, we look for the extensions in the form
\begin{equation}\label{IYe:2}
  \hat{Q}_m=Q_m+a_m(x_i,R)\partial_R.
\end{equation}

{\bf Definition 3}. Extension of a realisation of the Lie algebra L with the basis operators - first-order differential operators of the form (\ref{IYe:fodo}) is constructed of the same operators with more variables added.

To find scalar relative differential invariants, we need to add just one new variable, and to find only linear extensions.
 \begin{equation}\label{IYe:3}
  \hat{Q}_m=Q_m+a_m(x_i)R\partial_R.
\end{equation}

I used the lists of non-equivalent realisations of two-dimensional Lie algebras in \cite{{IYe:Maryna},{IYe:PBNL2003}}. Please see the additional references there.

We must note that the idea of equivalence/non-equivalence for the extended realisation accounts for classification of the RDI - the difference with classification just with respect to the local transformation will be seen on the example of the translation operators. Our classification of the linear extensions has in mind the following definition of equivalence of the relative differential invariants.

{\bf Definition 4}.
Two relative differential invariants of the algebra ${Q}$ are
called equivalent, if they are equivalent wrt transformations
from the equivalence group of the relative invariance
conditions.
\begin{equation}
\mathop{Q}\limits^l\!{}_m R(x,
u,\mathop{u}\limits_1,\ldots,\mathop{u}\limits_l)= \lambda_m(x,
u,\mathop{u}\limits_1,\ldots,\mathop{u}\limits_l)R. \nonumber
\end{equation}

We can also consider equivalence of pairs $(R,\lambda)$ of RDI with their respective multiplicators. It may be useful for practical purposes of description of invariant differential equations, as a linear combination of RDIs with the same multiplicator will also be an RDI and may be used to construct an  invariant differential equation.

The procedure for description of relative differential invariants proposed in \cite{IYe:IYeProcSPT2007}:

\begin{enumerate}

\itemsep=0pt
\item[1.] Construct Lie prolongations of the operators $Q_m$.

\item[2.] Write operators of extended action.

\item[3.] Classify realisations of the extended action up to transformations from the equivalence group of the invariance conditions.

\item[4.] Find a functional basis of ADI for the inequivalent realisations.

\item[5.] Construct RDI and ADI of the algebra $Q$ from absolute
invariants of operators of extended action by elimination of ancillary variables.

\end{enumerate}

Note that ancillary variables $R$ may enter the ADIs of the operators of the extended realisation as multipliers of the form $R^K, K\ne 0$ are some integers - the ADIs having the form $FR^K$, where $F$ are some functions of dependent and independent variables, and derivatives of the dependent variables of the relevant order, will produce the RDIs of the form $F$.

Let us remind properties of RDIs - a product of RDIs is also an RDI, an RDI in some non-zero degree will also be an RDI.

\section{Extensions of algebras of the translation operators}

We start from a seemingly very easy case, that is a one-dimensional Lie algebra. It should be considered anyway for the purpose of comprehensive presentation. It is a Lie algebra whose basis consists of one operator. Any one separate first-order differential operator obviously forms a Lie algebra, and is locally equivalent to a translation operator.

This case is also interesting as consideration of equivalence is somewhat different for the situations when we are interested in just equivalence with respect to the local transformations of variables, and classification with the purpose of classification of relative differential invariants.

Let us first construct a linear extension for the translation operator $\partial_x$.
We will look for it in the form
$$\hat{Q}=\partial_x + R(x)F\partial_F.$$

It is easy to check that an arbitrary $R(x)$ will satisfy the commutator criterion for this algebra.

Our task is to classify linear extensions of realisations - that is to list such linear extensions that cannot be transformed into one another by local transformations.

As to the standard classification up to equivalence with respect to the local transformations, we may use the Lie theorem on straightening out of vector fields as it was done e.g. in \cite{IYe:JMathPhys31-1995} (see \cite{IYe:Olver1}).

However, though the operator
\begin{equation}\label{IYe:tplusR}
\hat{Q}=\partial_x + a(x)R\partial_R
\end{equation}
\noindent
is certainly locally equivalent to $Q=\partial_x$,
for our purpose of the RDI classification we need to consider operators with a non-vanishing coefficient at $\partial_R$. It is easy to see that a non-extended $Q=\partial_x$ does not produce any RDIs, but the extended operator (\ref{IYe:tplusR}) gave us a RDI $R=\exp x$. This RDI, as well as other exponential RDIs of the translation operators, is not useful at all to describe invariant equations, but it should be listed if we aim at obtaining a comprehensive classification of RDIs.
So, the procedure of classification of RDIs requires classification of linear extended algebras with nonzero coefficients at $R\partial_R$.

Proper classification procedures are described e.g. in
\cite{IYe:OlverKamranEq} and in references therein, as well as in the papers on classification of realisations of low-dimensional Lie algebras of the first-order differential operators \cite{{IYe:Maryna},{IYe:PBNL2003}}.

Such classification in the case of operators considered in this section shall find algebras or single operators whose actions are not equivalent under local transformations of the following form:
\begin{gather}\nonumber
  \tilde{x}=\kappa(x,R), \qquad \tilde{R}=\phi(x,R) \\\nonumber
  \tilde{Q}_m= \tilde{a}_m(\tilde{x}) \partial_{\tilde{x}}+
   \tilde{b}_m(\tilde{x})\tilde{R} \partial_{\tilde{R}}
\end{gather}

Similar criteria would be also relevant for algebras involving more variables. However, for the purposes of this paper we can easily check equivalence or not equivalence using invariants of the relevant algebras or operators.

\section{Extensions of two-dimensional Lie algebras}
Following \cite{IYe:IYeProcSPT2007}, we will consider extensions of the Lie algebras
\begin{equation}\label{IYe:2d1}
\partial_x, \qquad x\partial_x;
\end{equation}
\begin{equation}\label{IYe:2d2}
\partial_x, \qquad y\partial_x;
\end{equation}
\begin{equation}\label{IYe:2d3}
\partial_x, \qquad x\partial_x +\partial_y,
\end{equation}
\noindent
excluding from our consideration here the algebras that consist only of the translation operators.

The commutator of operators (\ref{IYe:2d1}) is $\partial_x$.
We consider the extension
$$\partial_x+a(x,y)R \partial_R, \qquad x\partial_x+b(x,y)R \partial_R.$$
$b_x-xa_x=a$, then we  get determining conditions for the coefficients
$-xa=\phi(y)$; $a$ is arbitrary. We get
$b=xa(x,y)+\phi(y)$, and the general form of the extended operators
$$\partial_x+a(x,y)R \partial_R,\qquad x\partial_x+(xa(x,y)+\phi(y))R \partial_R.$$

We look for extensions of the algebra (\ref{IYe:2d2}) in the form
$$\partial_x+a(x,y)R \partial_R, \qquad y\partial_x+b(x,y)R \partial_R. $$
From the commutation relations of the algebra we get conditions on the coefficients of the extended operators
$b_x-ya_x$, then $b=ya(x,y)+\phi(y)$, and the extended basis operators of the algebra are as follows:
$$\partial_x+a(x,y)R \partial_R, \qquad y\partial_x+(ya(x,y)+\phi(y))R\partial_R. $$

We look for extensions of the algebra (\ref{IYe:2d3}) in the form
$$\partial_x+a(x,y)R \partial_R, \qquad x\partial_x +\partial_y+b(x,y)R \partial_R.$$
From the commutation relations of the algebra we get conditions on the coefficients of the extended operators
$b_x-xa_x-a_y=a$, then $b+xa=\Phi_y(x,y)$, $a=\Phi_x(x,y)$.
$b=\Phi_y(x,y)-x\Phi_x(x,y)$.

Extended basis operators of the algebra look as
$$\partial_x+\Phi_x(x,y)R \partial_R,\qquad x\partial_x+(\Phi_y(x,y)-x\Phi_x(x,y))R \partial_R.$$

\vspace{1ex}

{\small \hfill Table 1}
{\small
\begin{equation} \nonumber
\renewcommand{\arraystretch}{1.4}
\begin{tabular}{|l|l|p{5cm}|p{4.5cm}|}
\hline
&\hfil Basis Operators\hfil& General Extended Basis Operators \hfil&Inequivalent Extended Basis Operators \hfil \\ \hline
$\!1\!$&$\partial_x,
 x\partial_x;$&$\partial_x+a(x,y)R\partial_R$,\ \ \phantom{1111111111} \ \
 $x\partial_x+(xa(x,y)+\phi(y))R\partial_R;$& $\partial_x+R \partial_R$,\ \ \phantom{111111111111} \ \ $x(\partial_x+R \partial_R)+\epsilon R \partial_R;$ \\ \hline
$\!2\!$&$\partial_x, \ y\partial_x;$ & $\partial_x+a(x,y)R \partial_R$, \ \ \phantom{11111111111} \ \ $y\partial_x+(ya(x,y)+\phi(y))R\partial_R$; & $\partial_x+R \partial_R$, \ \ \phantom{1111111111111} \ \ $y\partial_x+(y+\epsilon)R\partial_R$;  \\ \hline
$\!3\!$&$\partial_x, \ x\partial_x +\partial_y;$& $\partial_x+\Phi_x(x,y)R \partial_R$, \ \ \phantom{11111111} \ \ $x\partial_x+(\Phi_y-x\Phi_x)R \partial_R.$& $\partial_x+\Phi_x(x,y)R \partial_R$,\ \ \phantom{11111111} \ \ $x\partial_x+(\Phi_y-x\Phi_x)R \partial_R.$ \hfil \\ \hline
\end{tabular}\end{equation}
}

In Table 1, $a(x,y)$, $\phi(y)$, $\Phi(x,y)$ are arbitrary sufficiently smooth functions; $\epsilon$ can take values $0$ or $1$.

\section{RDIs of two-dimensional Lie algebras}
Operators listed in Table 1 would allow calculation of zero-order relative differential invariants. Finding higher-order RDIs would require finding extensions of the prolongations of operators of the initial realisation to the relevant order. We would like to point out that here we need extensions of the relevant prolongations of the operators being considered - not prolongations of extensions.

Here we will look for the {\bf functional bases} of the RDI up to the second order of derivatives.

We find invariants for two independent variables $x$ and $y$, and one dependent variable $u$. Let us point out that finding differential invariants depends essentially from choice and assignment of dependent and independent variables.

{\small \hfill Table 2}
{\small
\begin{equation} \nonumber
\renewcommand{\arraystretch}{1.4}
\begin{tabular}{|l|p{4cm}|p{4cm}|p{4cm}|}
\hline
&Extended Second Prolongation \hfil& Functional Bases of ADI \hfil&Functional Bases of RDI \hfil \\ \hline
$\!1\!$&$\partial_x+R\partial_R$,\ \
 $x\partial_x-u_x\partial_{u_x}-
 u_{xx}\partial_{u_{xx}}-u_{xy}\partial_{u_{xy}}+R\partial_R;$&
 $y$, $u$, $u_y$, $u_{yy}$, $u_xR$, $u_{xx}R$, $u_{xy}R$; &$y$, $u$, $u_y$, $u_{yy}$, $u_x$, $u_{xx}$, $u_{xy}$; \\ \hline
$\!2\!$&$\partial_x+R\partial_R, \ y\partial_x-u_x\partial_{u_y}-
u_{xx}\partial_{u_{xy}}-u_{xy}\partial_{u_{yy}}
 +R\partial_R$; &$y$, $u$, $u_x$, $u_{xx}$, $\exp{\frac{u_y}{u_x}}R$,
 $u_xu_{xy}-u_yu_{xx}$,
 $u^2_{xy}-$ $2u_{xx}u_{yy}$; &$y$, $u$, $u_x$, $u_{xx}$, $\exp{u_y}$, $u_x u_{xy}-{u_y}u_{xx}$, $u^2_{xy}-$ $2u_{xx}u_{yy}$; \\ \hline
$\!3\!$& $\partial_x+R\partial_R$,\ \
 $x\partial_x+\partial_y-u_x\partial_{u_x}-
 u_{xx}\partial_{u_{xx}}-
 u_{xy}\partial_{u_{xy}}+R\partial_R$;
 &$\frac{\exp y}{R}$, $u$, $u_y$, $u_{yy}$, $u_xR$, $u_{xx}R$, $u_{xy}R$;
 & $\exp y$, $u$, $u_y$, $u_{yy}$, $u_x$, $u_{xx}$, $u_{xy}$.\\ \hline
\end{tabular}\end{equation}
}

For the first algebra in Table 2 only the last three RDIs in the list are proper RDIs. For the second one - only $\exp{u_y}$ is a proper RDI, and for the third algebra the proper RDIs in the list are
$\exp y,u_x, u_{xx}, u_{xy}$.

\section{General extensions of realisations: example of the Poincar\'e algebra for one space dimension}
Previously, many extensions were studied for many famous algebras of the mathematical physics without limitations of linearity and without any relation to finding relative differential invariants.

These nonlinear realisations were used to find their differential invariants and whence new equations invariant under these algebras.

Here I will remind in more detail only one my paper  \cite{IYe:IAYnonlin} on a nonlinear realisation of the Poincar\'e algebra $P(1,2)$. However, here it is sufficient to refer for an illustration a simpler case of $P(1,1)$ and its nonlinear realisation found in \cite{IYe:JMathPhys31-1995}.

We will illustrate the difference in the problem of classification of the general extensions for Lie algebras and of the extensions with the purpose of the RDI classification on the example of the Poincar\'e algebra with one space dimension, two independent variables $t$ and $x$ and one dependent variable $u$:
\begin{equation}\label{IYe:poin1}
\partial_t, \qquad \partial_x, \qquad J=t\partial_x + x\partial_t.
\end{equation}

A functional basis of the second-order ADIs for this realisation can be written as follows \cite{IYe:JMathPhys31-1995}:
 \begin{gather}\label{IYe:ADIpoin1}
I_1=u, \qquad I_2=u^2_t-u^2_x, \qquad I_3=u_{tt}-u_{xx}, \\ \nonumber
I_4=(u_t-u_x)^2(u_{tt}+2u_{tx}+u_{xx}), \quad
I_5=(u_t+u_x)^2(u_{tt}-2u_{tx}+u_{xx}).
\end{gather}

Setting $I_4$, $I_5$ to zero, the authors actually obtained expressions that are relative differential invariants
\begin{gather}\label{IYe:1RDIpoin1}
AR_1=u_t-u_x, \qquad
AR_2=u_t+u_x, \\
AR_3=u_{tt}+2u_{tx}+u_{xx}, \ \nonumber
AR_4=u_{tt}-2u_{tx}+u_{xx},
\end{gather}

\noindent
by means of listing invariant equations of the type
$AR_i=0$, but did not mention the concept of a relative differential invariant, and did not give any statements on full classification of such invariants.

Let us look at the extension of the standard realisation of the Poincar\'e algebra (\ref{IYe:poin1}) constructed with the aim of classification of RDIs.
\begin{gather}\label{IYe:poin1exgen}
 \partial_t + a(t,x)R\partial_R, \qquad\
 \partial_x+b(t,x)R\partial_R, \\ \nonumber
 J=t\partial_x + x\partial_t +c(t,x)R\partial_R.
\end{gather}

From the commutation relations
\begin{gather} \label{IYe:commrel1}
[P_t, P_x]=0, \qquad
[P_t,J]= P_x, \qquad
[P_x,J]=P_t
\end{gather}
\noindent
we get conditions on the functions $a(t,x)$, $b(t,x)$,
$c(t,x)$:
\begin{gather}\label{IYe:cond-poin1}
 a_x=b_t, \qquad
 c_t-ta_x-xa_t=b, \\ \nonumber
 c_x-tb_x-xb_t=a,
\end{gather}
\noindent
whence
$$a(t,x)=\Phi_t, \qquad b(t,x)=\Phi_x, \qquad c(t,x)=t\Phi_x+x\Phi_t+C,$$
\noindent
where $\Phi=\Phi(t,x)$ is an arbitrary sufficiently smooth function of its arguments, and $C=const$.

Up to local equivalence and on condition of non-zero coefficients at $R\partial_R$, we obtain the following realisation
\begin{gather}\label{IYe:poin1ex-ne}
 \partial_t + R \partial_R, \qquad
 \partial_x+R \partial_R, \\ \nonumber
 J=t(\partial_x+ R \partial_R) +
 x(\partial_t + R \partial_R) +\epsilon R \partial_R,
\end{gather}
\noindent
where
$ \epsilon$ is equal to 0 or 1.

Operators (\ref{IYe:poin1ex-ne})  with $\epsilon=0$ give give from its functional  basis of ADIs $R^{-1}\exp t$,
$R^{-1}\exp x$ a set of RDIs
$\exp t$, $\exp x$.

We can find the extended prolongation of realisation
(\ref{IYe:poin1}) using the commutation relations similarly, and obtain:
\begin{gather}\nonumber
 \partial_t + R\partial_R, \
 \partial_x+R\partial_R, \\
 \label{IYe:poin1ex2-ne}
 J=t(\partial_x+R\partial_R) + x(\partial_t + R\partial_R) - u_t\partial_{u_x}- u_x\partial_{u_t}
 - \\ \nonumber
 u_{tt}\partial_{u_{xt}}-2u_{xt}(\partial_{u_{xx}} +
\partial_{u_{tt}})-u_{xx}\partial_{u_{xt}}+\epsilon R\partial_R.
\end{gather}

Operators (\ref{IYe:poin1ex2-ne}) with $\epsilon=1$ give from its functional  basis of first-order ADIs
\begin{equation} \label{IYe:ADI1poin1}
(u_t+u_x)R^{-1}, \ \ (u_t-u_x)R
\end{equation}
\noindent
first-order RDIs for the algebra (\ref{IYe:poin1})
\begin{equation}\nonumber
u_t+u_x, \ \ u_t-u_x.
\end{equation}

The determining equation for the second-order ADIs of the form $F=F(u_{tt},u_{xt},u_{xt})$
of (\ref{IYe:poin1}),  will look as follows:
\begin{equation}\nonumber
2u_{xt}(F_{u_{xx}} +
F_{u_{tt}})+(u_{xx}+u_{tt})F_{u_{xt}}+RF_R=0.
\end{equation}
The resulting second-order ADIs are
\begin{equation}\label{IYe:ADI2poin1}
u_{tt}-u_{xx}, \ (u_{tt}+2u_{xt}+u_{xx})R^{-2}, \
(u_{tt}-2u_{xt}+u_{xx})R^2.
\end{equation}
In the same way we obtain two inequivalent proper second-order RDIs:
\begin{equation}\nonumber
u_{tt}+2u_{xt}+u_{xx}, \ \ u_{tt}-2u_{xt}+u_{xx}.
\end{equation}

An invariant $u_{tt}-u_{xx}$ is an ADI of the realisation (\ref{IYe:poin1}), so we do not include it into the list of RDIs.

It is easy to see that products of invariants in the relevant degrees from the lists (\ref{IYe:ADI1poin1}), (\ref{IYe:ADI2poin1}) to eliminate ancillary variables $R$ will give
absolute invariants from the list (\ref{IYe:ADIpoin1}). So
 to describe all non-equivalent RDIs up to the second order of the realisation being considered it is sufficient to take only zero-and first-order RDIs in addition to the list of ADIs.

The general extensions of the realisation (\ref{IYe:poin1}) were studied in \cite{IYe:JMathPhys31-1995}. The authors of \cite{IYe:JMathPhys31-1995} found a new extended realisation
 \begin{equation}\label{IYe:poin2}
 \partial_t, \ \partial_x, \ \ J=t\partial_x + x\partial_t +u\partial_u
\end{equation}

\noindent
that is not locally equivalent to the realisation (\ref{IYe:poin1}), and found absolute differential invariants of (\ref{IYe:poin2}) up to the second order.

\begin{gather}\nonumber
I_1=u_t+u_x, \ I_2=(u_t-u_x)u^{-2}, \
I_3=(u_{tt}-u_{xx})u^{-1}, \\ \nonumber
I_4=(u_{tt}+2u_{tx}+u_{xx})u, \ \
I_5=(u_{tt}-2u_{tx}+u_{xx})u^{-3}.
\end{gather}

To obtain a comprehensive classification of RDIs for this extended realisation, we need to extend it further and to consider the realisation
 \begin{gather}\nonumber
 P_t=\partial_t + a(t,x,u)R\partial_R,\qquad
 P_x=\partial_x+b(t,x,u)R\partial_R, \\ \nonumber
 J=t\partial_x + x\partial_t +u\partial_u +c(t,x,u)R\partial_R.
\end{gather}

From the commutation relations
\begin{gather} \nonumber
[P_t, P_x]=0, \qquad
[P_t,J]= P_x, \qquad
[P_x,J]=P_t
\end{gather}

we get conditions on the functions $a(t,x,u)$, $b(t,x,u)$,
$c(t,x,u)$:
\begin{gather} \nonumber
 a_x=b_t, \qquad
 c_t-ta_x-xa_t-ua_u=b, \\ \nonumber
 c_x-tb_x-xb_t-ub_u=a,
\end{gather}
\noindent
whence
$$a(t,x,u)=\Phi_t, \qquad b(t,x,u)=\Phi_x, \qquad c(t,x,u)=t\Phi_x+x\Phi_t+u\Phi_u+C,$$
\noindent
where $\Phi=\Phi(t,x,u)$ is an arbitrary sufficiently smooth function of its arguments, and $C=const$.

Up to local equivalence and on condition of non-zero coefficients at $R\partial_R$, we obtain the following realisation
\begin{gather}\label{IYe:poin1ex-ne1}
 \partial_t + R \partial_R, \qquad
 \partial_x+R \partial_R, \\ \nonumber
 J=t(\partial_x+ R \partial_R) +
 x(\partial_t + R \partial_R) + u\partial_u+\epsilon R \partial_R,
\end{gather}
\noindent
where
$ \epsilon$ is equal to 0 or 1.

Operators (\ref{IYe:poin1ex-ne1})  with $\epsilon=0$ give from its functional  basis of ADIs
$$R^{-1}\exp t, \qquad
R^{-1}\exp x$$
\noindent
a set of RDIs
$$\exp t, \qquad \exp x.$$

We can find an extended prolongation of realisation
(\ref{IYe:poin1}) using the commutation relations similarly, and obtain:
\begin{gather}\nonumber
 \partial_t + R\partial_R, \
 \partial_x+R\partial_R, \\ \nonumber
 \label{poin1ex2-ne1}
 J=t(\partial_x+R\partial_R) + x(\partial_t + R\partial_R) + u\partial_u +u_x\partial_{u_x}
 +u_t\partial_{u_t}\\ \nonumber
 +u_{xx}\partial_{u_{xx}}
 +2u_{xt}\partial_{u_{xt}}+u_{tt}\partial_{u_{t}} - u_t\partial_{u_x}- u_x\partial_{u_t}
 - \\ \nonumber
 u_{tt}\partial_{u_{xt}}-2u_{xt}(\partial_{u_{xx}} +
\partial_{u_{tt}})-u_{xx}\partial_{u_{xt}}+\epsilon R\partial_R.
\end{gather}

We can take relative differential invariants
as follows:
\begin{gather}\nonumber
IR_1= u, \ I_2=u_t+u_x, \ IR_3=u_t-u_x, \
IR_4=u_{tt}-u_{xx}, \\ \nonumber
IR_5=u_{tt}+2u_{tx}+u_{xx}, \ \
IR_6=u_{tt}-2u_{tx}+u_{xx}.
\end{gather}

\section{Further work}
The idea of extended realisations for which we find absolute dif\/ferential invariants
provides for the possibility to use all the extensive theory, available in the literature for such absolute dif\/ferential invariants, in particular, ability to construct operators of invariant differentiation and fundamental bases of ADIs, enabling description of invariants of any order. See e.g. starting from the classical papers \cite{IYe:LieDI}, \cite{IYe:Tresse}, \cite{IYe:Spencer} and  \cite{IYe:Glenn1928}; as well as more recent books \cite{IYe:Olver2} and \cite{IYe:Olver3}, and papers \cite{IYe:Beffa-Sanders}, \cite{IYe:BeffaRDI}, \cite{IYe:Boutin}.

It would be interesting to study RDIs for more algebras, including physically interesting ones, and to look for new nonlinear realisations of such algebras, continuing work in e.g. \cite{IYe:JMathPhys31-1995}, \cite{IYe:IAYnonlin}, \cite{IYe:FZhL94}, \cite{IYe:NestVGalilei2016}

Relative differential invariants are needed for complete group classification of PDE. We provided a full description of ADIs for the Poincar\'e algebra and arbitrary number of scalar functions and arbitrary number of independent variables, but a similar description for vector functions is still not available. We believe that utilisation of extended realisations may be useful for this purpose. Though, some ideas in this respect for low numbers of independent variables are given in \cite{IYe:IYePreprcomplfields}.

\section{Acknowledgements}
The first and foremost my acknowledgements go to the
Armed Forces of Ukraine due to whom I am alive. Please remember that Russia is an aggressor country and still plans to kill all Ukrainians not going to be their zombies, and all 100 percent of Russian scientists contribute to killings despite words about peace from a tiny portion of them.

I would like also to thank the Institute of Mathematics of the Polish Academy of Sciences for their hospitality and grant support, to the National Academy of Sciences of the USA and the National Centre of Science of Poland for their grant support.

I would like also to mention my good memories and appreciation of my visit to the Centre de recherches math\'ematiques in Montreal by invitation of late Ji\v{r}\`i Patera where I had fruitful discussions with Ji\v{r}\`i Patera and Pavel Winternitz, and I also was able to study literature on differential invariants, and, in particular, find papers by Oliver Glenn.

Research was supported by Narodowe Centrum Nauki, grant number 2017/26/A/ST1/00189.

\end{document}